\documentclass[twocolumn,nofootinbib,10pt]{revtex4-2}

\usepackage{amsmath}
\usepackage{amssymb}
\usepackage{wasysym}
\usepackage{graphicx}
\usepackage{color,soul}
\usepackage{physics}
\usepackage{siunitx}
\usepackage{dsfont}
\usepackage{float}
\usepackage{ulem}
\usepackage[english]{babel}
\usepackage{blindtext}
\usepackage[english,nomargin,inline,marginclue,draft]{fixme}
\pdfpageheight\paperheight
\pdfpagewidth\paperwidth

\usepackage[colorlinks,linkcolor=blue,anchorcolor=blue,citecolor=blue,urlcolor=blue]{hyperref}

\fxusetheme{colorsig}
\FXRegisterAuthor{cg}{acg}{CG}
\FXRegisterAuthor{th}{ath}{\color{blue}TH}
\FXRegisterAuthor{ib}{aib}{\color{red}IB}
\FXRegisterAuthor{sh}{ash}{\color{cyan}SH}
\FXRegisterAuthor{db}{adb}{\color{green}DB} 
\FXRegisterAuthor{ps}{aps}{PS}
\makeatletter
\renewcommand*\FXLayoutInline[3]{%
  {\@fxuseface{inline}\ignorespaces{\color{fx#1}[#3: #2]}}}
\makeatother

\long\def\symbolfootnote[#1]#2{\begingroup%
\def\thefootnote{\fnsymbol{footnote}}\footnotetext[#1]{#2}\endgroup}

\def\nobreakbefore{%
  \relax\ifvmode\else
    \ifhmode
      \ifdim\lastskip > 0pt\relax
        \unskip\nobreakspace
      \else 
        \nobreakspace
      \fi
    \fi
  \fi
}
\let\oldcite\cite
\renewcommand\cite{\nobreakbefore\oldcite}

\begin{document}
\title{Ergodicity breaking from Rydberg clusters in a driven-dissipative many-body system}

\author{Dong-Sheng Ding$^{1,2,\textcolor{blue}{\ast},\textcolor{blue}{\dag}}$}
\author{Zhengyang Bai$^{3,\textcolor{blue}{\ast}}$}
\author{Zong-Kai Liu$^{1,2,\textcolor{blue}{\ast}}$}
\author{Bao-Sen Shi$^{1,2,\textcolor{blue}{\ddagger}}$}
\author{Guang-Can Guo$^{1,2}$}
\author{Weibin Li$^{4,\textcolor{blue}{\mathsection}}$}
\author{C. Stuart Adams$^{5,\textcolor{blue}{\mathparagraph}}$}

\affiliation{$^1$Key Laboratory of Quantum Information, University of Science and Technology of China, Hefei, Anhui 230026, China.}
\affiliation{$^2$Synergetic Innovation Center of Quantum Information and Quantum Physics, University of Science and Technology of China, Hefei, Anhui 230026, China.}
\affiliation{$^3$State Key Laboratory of Precision Spectroscopy, East China Normal University, Shanghai 200062, China.}
\affiliation{$^4$School of Physics and Astronomy, and Centre for the Mathematics and Theoretical Physics of Quantum Non-equilibrium Systems,
University of Nottingham, Nottingham, NG7 2RD, United Kingdom.}
\affiliation{$^5$Department of Physics, Joint Quantum Centre (JQC) Durham-Newcastle, Durham University, South Road, Durham DH1 3LE, United Kingdom.}
\date{\today}

\symbolfootnote[1]{These authors contributed equally to this work.}
\symbolfootnote[2]{dds@ustc.edu.cn}
\symbolfootnote[3]{drshi@ustc.edu.cn}
\symbolfootnote[4]{weibin.li@nottingham.ac.uk}
\symbolfootnote[5]{c.s.adams@durham.ac.uk}

\maketitle

\textbf{It is challenging to probe ergodicity breaking trends of a quantum many-body system when dissipation inevitably damages quantum coherence originated from coherent coupling and dispersive two-body interactions. Rydberg atoms provide a test bed to detect emergent exotic many-body phases and non-ergodic dynamics where the strong Rydberg atom interaction competes with and overtakes dissipative effects even at room temperature. Here we report experimental evidence of a transition from ergodic towards ergodic breaking dynamics in  driven-dissipative Rydberg atomic gases. The broken ergodicity is featured by the long-time phase oscillation, which is attributed from the  formation of  Rydberg excitation clusters in limit cycle phases. The broken symmetry in the limit cycle is a direct manifestation of many-body interactions, which is verified by tuning atomic densities in our experiment. The reported result reveals that Rydberg many-body systems are a promising candidate to probe ergodicity breaking dynamics, such as limit cycles, and enable the benchmark of non-equilibrium phase transition.
}


Many-body systems typically relax to  equilibrium due to ergodicity such that observable becomes invariant with time~\citep{srednicki1994chaos,srednicki1999approach,reichl_modern_2016}. Often, the equilibration is so robust such that the observable quickly seeks new fixed points in phase space after quenching control parameters, which is driven by Boltzmann’s ergodic hypothesis \citep{boltzmann1887ueber}. Exceptions, i.e. broken ergodicity due to symmetry breaking~\citep{venkataraman_beyond_1989,kinoshita2006quantum}, have been extensively explored in integrable~\cite{baxter_exactly_2008} and many-body localized systems~\cite{nandkishore_many-body_2015,abanin_colloquium_2019}.  A recent example is the quantum many-body scars~\cite{turner2018weak} of strongly interacting Rydberg atoms trapped in optical arrays~\cite{browaeys_many-body_2020}, where  non-ergodic many-body dynamics take place in a constrained sub-Hilbert space \citep{labuhn2016tunable,Orthogonal_Zhao_2022}. This leads to coherent revivals of the $\mathbb{Z}_2$ state~\citep{choi2019emergent} largely due to strong Rydberg atom interactions~\cite{lukin2001dipole,schauss2012observation,bernien2017probing, bluvstein_controlling_2021}.
In the presence of dissipation it is common that quantum many-body coherence and entanglement are eliminated, leading to stationary states  in long-time limit. The interplay between strong Rydberg interactions and dissipation, on the other hand, results to exotic non-equilibrium phenomena such as collective quantum jumps
\citep{lee2012collective}, and phase transitions \citep{lee2011antiferromagnetic,Liquid_Igor_2012,qian2013quantum,marcuzzi2014universal,PhysRevLett.113.0230062014,
weimer2015variational,levi2016quantum,PhysRevA.96.0416022017,dissipative2022time}. Optical bistability~\cite{carr2013nonequilibrium}, and self-organized criticality \citep{Ding_Phase_2020,Signatures_Helmrich_2020} have been demonstrated experimentally.

\begin{figure*}[t]
\includegraphics[width=1.9\columnwidth]{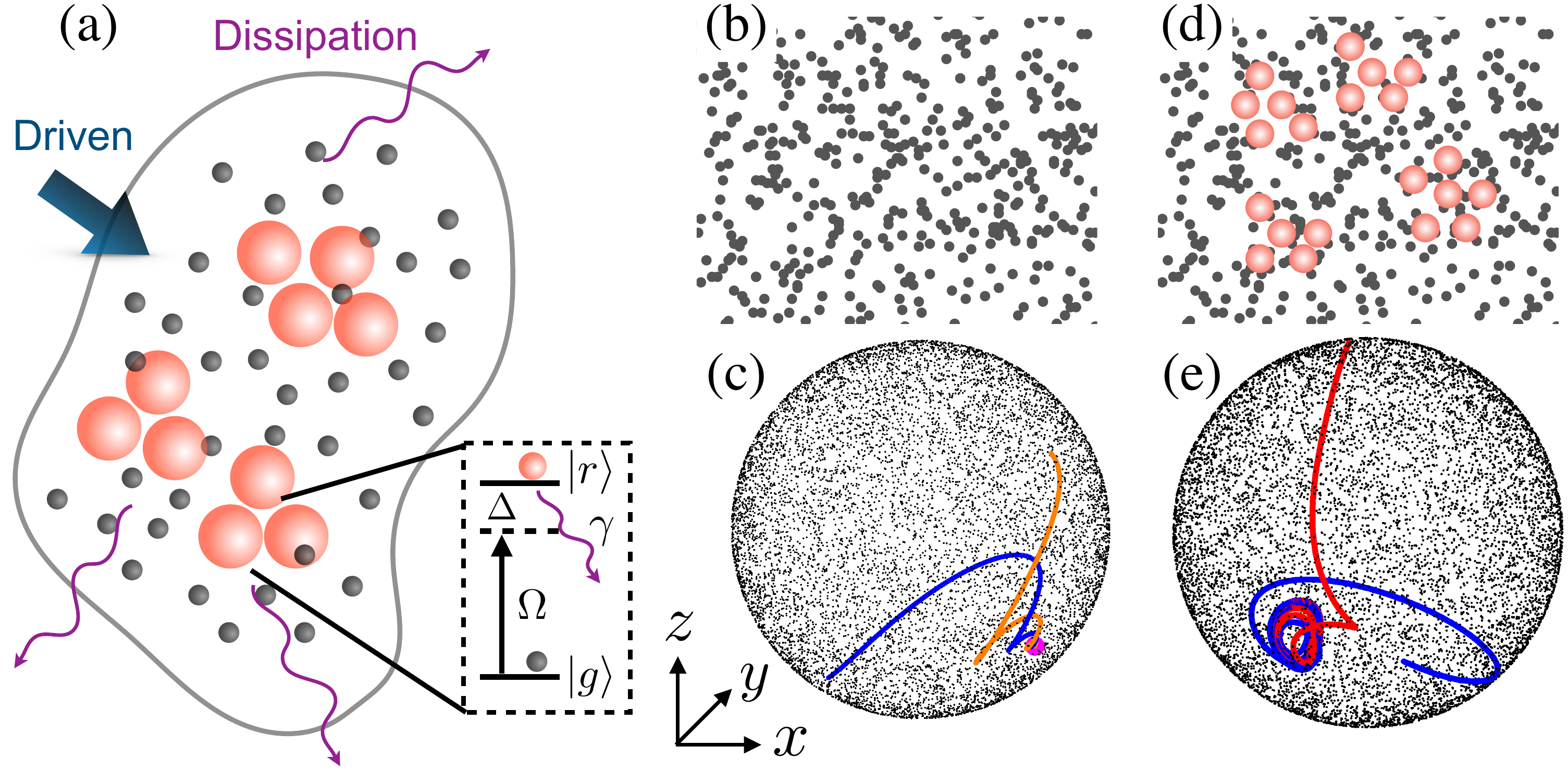}\caption{\textbf{Ergodic and non-ergodic phases in a driven-dissipative Rydberg gas.} (a) Atoms are laser excited from the ground state $\left|g\right\rangle $ (black sphere) to  Rydberg state  $\left|r\right\rangle $ (red, large sphere)  with detuning $\Delta$, Rabi frequency $\Omega$ and decay rate $\Gamma$. Due to the strong Rydberg interaction, Rydberg excitations are separated in space, where the minimal distance is determined by the blockade radius. (b) Ergodic phase. When the Rydberg atom interaction is weak, the dissipation leads to a single stationary state (fixed point). As shown in (c) any initial states on or in the Bloch sphere will decay to the fixed point (magenta dot). Blue and orange curves depict trajectories of atoms from two different initial conditions. (d) Non-ergodic phase. When the interaction is strong, closely packed Rydberg atoms form active clusters, as highlighted in the figure, leading to non-stationary dynamics. In phase space, the corresponding trajectories oscillate persistently, building up a limit cycle (e). }
\label{setup}
\end{figure*}


Here we report observation of non-ergodic many-body dynamics in a thermal gas of strongly interacting Rydberg atoms. This setting, as depicted in Fig.~\ref{setup}(a), is a dissipative
 many-body system of effective two-level atoms (ground and Rydberg states $|g\rangle$ and $|r\rangle$). Coherent laser coupling and strong long-ranged interactions in Rydberg state $|r\rangle$ compete with dissipation (Doppler and collisional effects, electronic state decay etc). A non-equilibrium phase transition is identified by quenching the detuning $\Delta$ provided that the Rabi frequency $\Omega$ is above a critical value, in which a bifurcation between an ergodic (E) and weakly non-ergodic (NE) phase appears. The E-phase corresponds to a weak interaction regime where the distribution of Rydberg atoms is homogeneous [Fig.~\ref{setup}(b)]. Irrelevant to initial states, excitations of all atoms end at an identical fixed point in phase space [Fig.~\ref{setup}(c)]. The NE phase features a non-trivial population revival when the laser detuning is near resonant. Long-time many-body coherence, in the order of milliseconds, is observed in the NE phase, where Rydberg population oscillations persist for a period much longer than any time scales of the relevant dissipation and laser Rabi frequency. Our analysis suggests that the broken ergodicity is induced by clustering of strongly interacting Rydberg atoms in \textit{free space}~\cite{garttner_dynamic_2013,lesanovsky_out--equilibrium_2014,urvoy_strongly_2015, letscher_bistability_2017}, where many-body dynamics are synchronized and form the oscillatory phase, i.e. a limit cycle [Figs.~\ref{setup}(d) and (e)]~\cite{Lee_Antiferromagnetic_2011}. The non-equilibrium dynamics is measured non-destructively through electromagnetically-induced transparency (EIT).  Due to the unprecedented level of controllability, thermal Rydberg atom vapor gases provide a platform to explore and probe non-ergodicity of matter in addition to the Rydberg array simulator~\cite{bernien2017probing, bluvstein_controlling_2021}.

\begin{figure*}
\includegraphics[width=2\columnwidth]{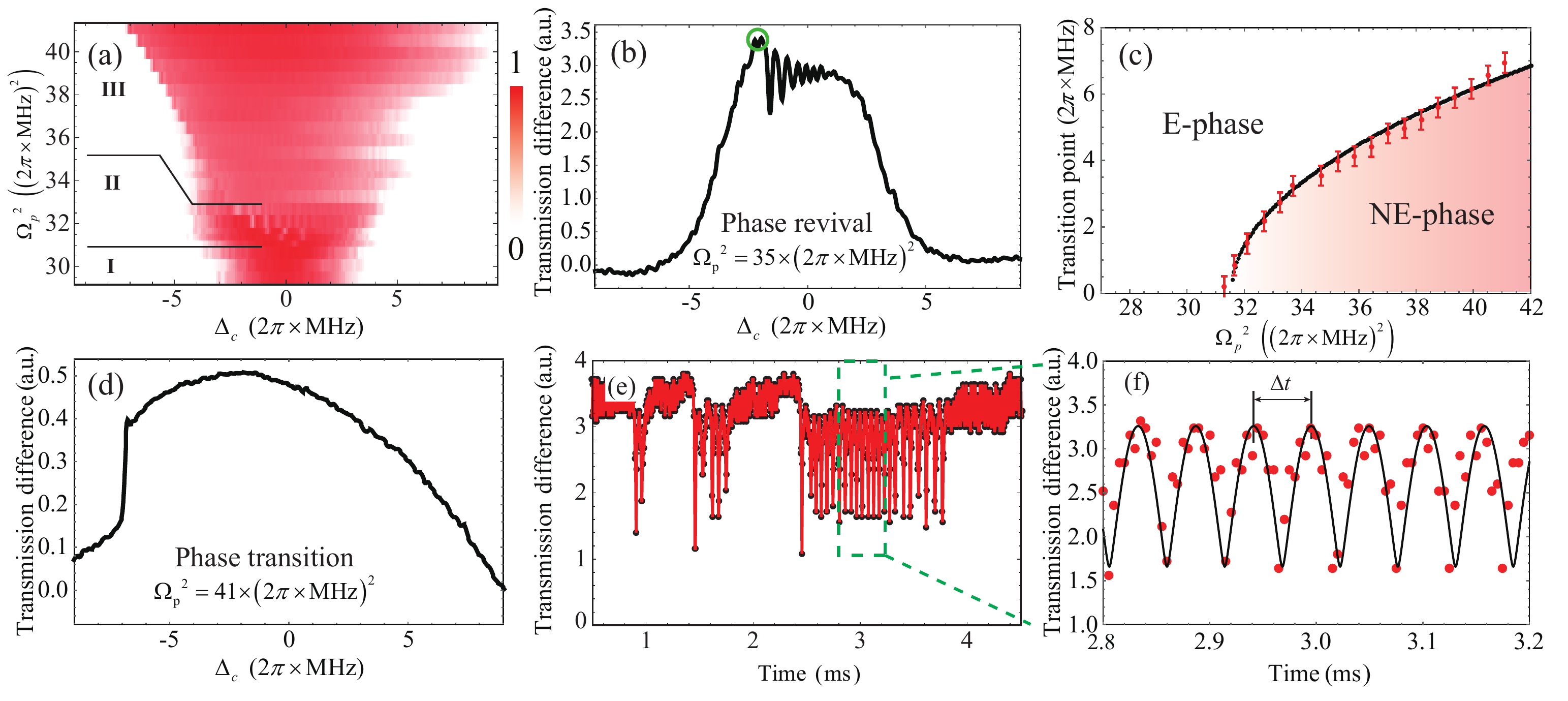}\caption{\textbf{Non-equilibrium phase diagram and ergodicity breaking transition.}  (a) The transmission spectrum as a function of $\Omega_{p}^{2}$ and $\Delta_{c}$ with the sweep rate  $2\pi\times6$ MHz/ms. Three regions \textbf{I}, \textbf{II} and \textbf{III} mark the ergodic, ergodicity breaking and strongly interacting phases, respectively. (b) When sweeping the detuning, transmission grows rapidly when close to the resonance, and the oscillates around the resonance. The maximal population (marked by the circle) indicates the critical population where the many-body system changes from ergodic to non-ergodic phase. In the latter case, the transmission does not relax to equilibration. (c) The phase transition point characterized by the parameters $|\Delta_{c}|$ and $\Omega_{p}^{2}$. The transition points are extracted from the first jump of phase revival regime of panel (a) as labeled in (b). The error bars are statistics according to the experimental fluctuations. (d) In the strongly interacting phase, the transmission increases sharply at the critical point. Before and after the rapid change, the transmission changes smoothly when sweeping the detuning. (e) The observed time flow of the many-body non-ergodic state when stopping sweep $\Delta_{c}$ in the vicinity of the first jump, i.e., $\Delta_{c} \sim -1.5\times 2\pi$ MHz. Long time phase oscillations are found in the experiment. (f) Enlarged view of the region labelled in (e). The black curve is the fitting curve with function $\sqrt{1.64+1.60 \cos{(1.88-116t)}}+1.46$. The period of collapse and revival is $\Delta t \approx 0.053$ms.}
\label{phase diagram}
\end{figure*}

\textbf{The Experiment.} We prepare Rubidium-85 atom gases above room temperature (typically $45^{\circ} \text{C}$) with density around $9.0\times10^{10} \text{cm}^{-3}$. The atom is excited from ground state $|g\rangle$ to Rydberg state $|r\rangle$ through EIT, i.e. from ground state $|g\rangle=\left| 5S_{1/2},F=2 \right\rangle$ to intermediate state $|e\rangle=\left| 5P_{1/2},F'=3 \right\rangle$ and then to Rydberg-state $|r\rangle=\left| 51D_{3/2} \right\rangle$ by a probe and coupling light field, respectively. The Rabi frequency and detuning of the probe and coupling light are denoted by $\Omega_{p}$ and $\Delta_{c}$. Both can be time dependent in our experiment.  Here a 795 nm laser is split into a probe beam and an identical reference beam by a beam displacer, which are both propagating in parallel through a heated Rb cell ($10\,\mathrm{cm}$ long). The two-photon excitation with counter-propagating beams ensures that a narrow, low velocity class of atoms are excited to Rydberg states due to velocity selections~\cite{tanasittikosol_subnatural_2012}. The transmission signal of the probe beam is detected on a differencing photodetector as a transmission difference~\cite{carr2013nonequilibrium}. Rydberg atoms vary for different realizations in thermal gases, such that conventional schemes based on electrically ionizing Rydberg atoms can not be used here~\cite{Potvliege_2006, Geier_2021}. Through EIT, however, the transmission signal (proportional to Rydberg atom populations) is measured continuously and dynamically while not demolishing atomic states; see Section I of \textbf{Supplementary Materials (SM)} for details.



\textbf{Ergodic-breaking phase transition.} We probe the Rydberg population by changing the laser detuning $\Delta_c$ linearly from red to blue side while the probe laser is on resonance. Typically $\Delta_c$ is quenched at a rate $2\pi\times 6$ MHz/ms. By raising the probe intensity, transmission against $\Omega_{p}^{2}$ and $\Delta_{c}$ is recorded and shown in Fig.~\ref{phase diagram} (a).  The transmission is weak  if the laser is away from the resonance. Close to the resonance, we observe the transmission exhibits distinctive features depending on $\Omega_p$.

When the probe laser is weak \textbf{$\Omega_{p}^{2}<30.9$} $(2\pi\times\mathrm{MHz)^{2}}$, transmission adiabatically follows the detuning $\Delta_c$, leading to a smooth spectrum [region \textbf{I} in Fig.~\ref{phase diagram} (a)]. This is a dissipation dominant phase. Increasing $|\Omega_p|$, a startling difference is that the transmission spectrum oscillates with increasing detuning, marked by region \textbf{II} in Fig.~\ref{phase diagram} (a). Specifically the population of Rydberg atoms bifurcates from the E- to  NE-phases, which takes place when $\Omega_{p}^{2}>30.9$ $(2\pi\times\mathrm{MHz)^{2}}$. A typical example with $\Omega_p^2 = 35 (2\pi\times \text{MHz})^2$ is shown in Fig.~\ref{phase diagram}(b), where a sudden jump is found when we increase $\Delta_c$. After passing the jump, the spectrum oscillates and decreases gradually when further increasing $\Delta_c$. Such nontrivial spectrum profile is unique in region \textbf{II}.


\begin{figure*}[htbp]
	\includegraphics[width=2\columnwidth]{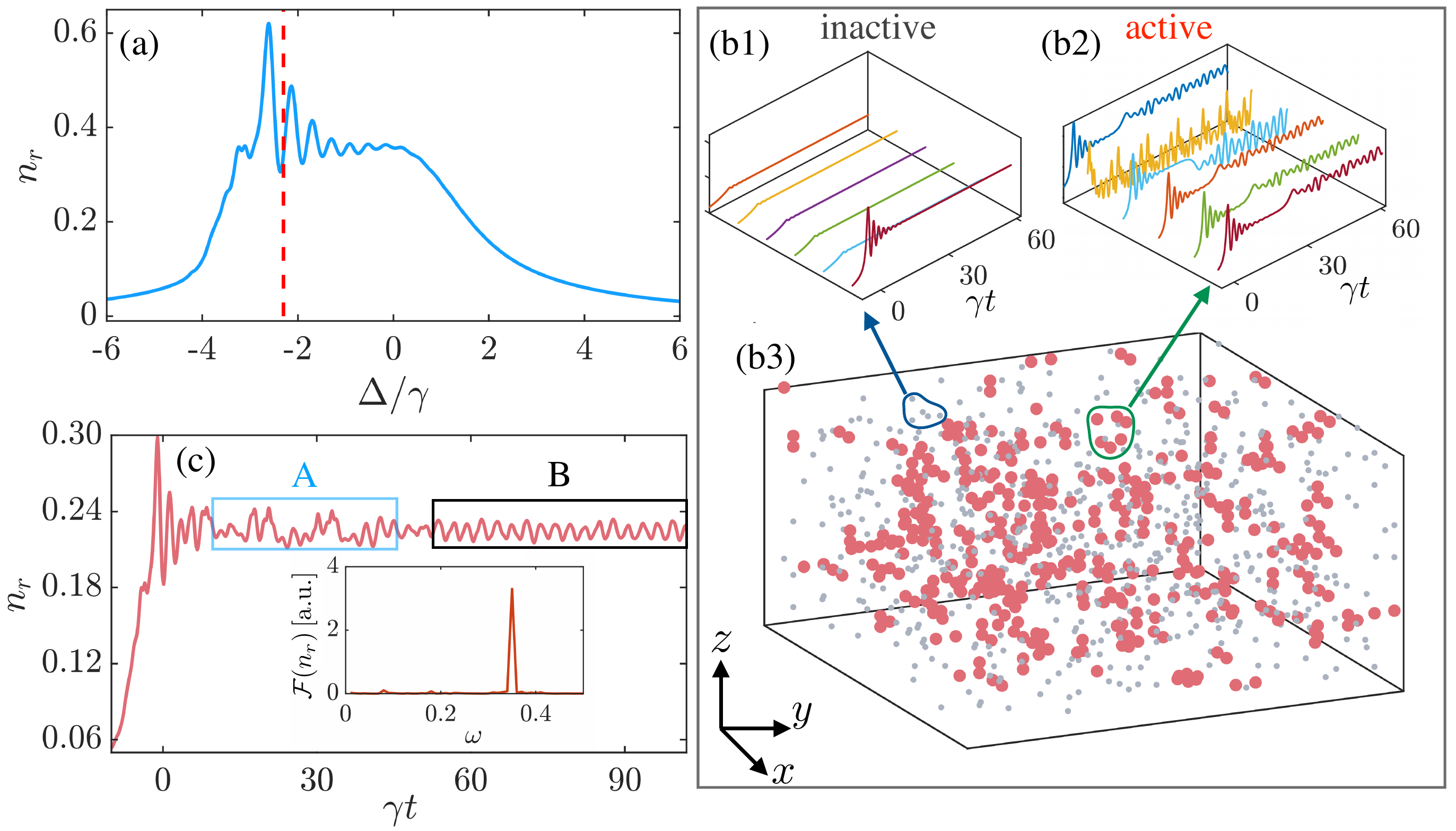}\caption{\textbf{Rydberg clusters and synchronized oscillations.}  (a) Rydberg population $n_r$ as a function of $\Delta$. At $t=0$, $\Delta(0) = -10\gamma$ and atoms are in the ground state. When $\Delta$ approaches to the resonance, Rydberg population grows and starts to oscillate. The vertical dashed line indicates the first jump point, which is observed experimentally. In a different scenario, the detuning is frozen at the first jump. It is found that most of atoms become dynamically inactive, where the population relaxes to equilibration rapidly (b1). Clusters of atoms, on the other hand, are active persistently (b2). In panel (b3) distributions of atoms in the ensemble are shown. The active and inactive atoms are denoted with gray and red dots. (c) Dynamical oscillations of the active atoms are synchronized after a transient period A. When the dynamics is synchronized (see region B for a finite duration), oscillations of atoms are centered at a single frequency (inset).}
	\label{against scan rate}
\end{figure*}

 The non-equilibrium phase transition occurs when the population of Rydberg atoms increases from $n_{r}<n_{r,c}$ (E-phase) to $n_{r}>n_{r,c}$ (NE-phase), where $n_{r,c}$ is a critical population obtained from the experiment. We plot a phase diagram in Fig.~\ref{phase diagram}(c) by recording the critical point at which the first jump appears from increasingly $\Delta_{c}$. The critical  point $\Delta_{c,c}$ is found to be
\begin{eqnarray}
	\Delta_{c,c}=a[1-(\Omega_p/\Omega_{p,c})^2]^b
\end{eqnarray}
where $\Omega_{p,c} = 2\pi \times 5.92$ MHz is the critical Rabi frequency of the probe beam, $b=0.61\pm0.04$ is the fitted critical exponent. This is consistent with
the prediction of the function $n_r[\Delta_{c},\Omega_{p}]-n_{r,c}=0$, where $n_r[\Delta_{c},\Omega_{p}]$
is the population of the steady state calculated from the  master equation. When $\Omega_p^2$ is large (region $\textbf{III}$ in Fig.~\ref{phase diagram} (a)), the transmission increases sharply to the maximal value, and then changes smoothly when increasing $\Delta_c$, shown in Fig.\ref{phase diagram} (d). In a lattice, this corresponds to the antiferromagnetic phase~\cite{Lee_Antiferromagnetic_2011}. As our theoretical analysis shown in the \textbf{SM}, here atoms end at different final states, while the overall transmission is stationary.  

To further understand the dynamics, we stop sweeping $\Delta_{c}$ near resonance and measure the evolution of the transmission. We observe a time flow of many-body state collapsing and revival periodically as given in Figs.~\ref{phase diagram}(e) and (f). The oscillation has a period of $\Delta T\approx 0.053$ ms. More importantly, it has very long lifetimes more than 1 ms, much longer than other characteristic time scales in the thermal Rydberg atom system.

\begin{figure*}[t]
	\includegraphics[width=2\columnwidth]{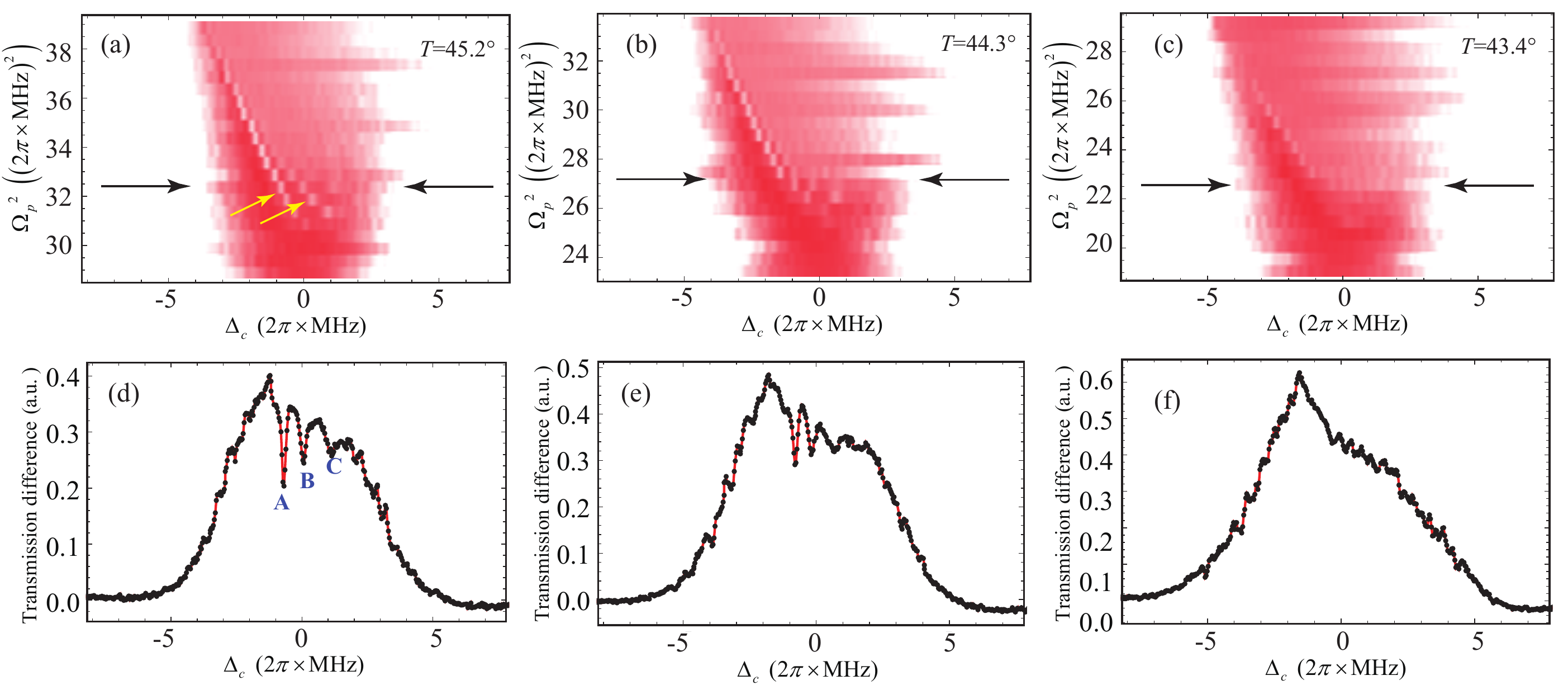}\caption{\textbf{Density-dependent phase diagram. }(a-c) The measured color-map of the probe transmission against temperatures from $T=45.2^{\circ} \text{C},44.3^{\circ}\text{C},43.4^{\circ}\text{C}$, corresponding to the atomic densities of $9.52,8.78,8.11\times10^{10}$
		$\mathrm{cm^{-3}}$. The yellow lines marked in (a) show obvious character of oscillations. (d-f) The decreased oscillated probe transmission spectra under lowering temperature. The data in (d-f) are the correspondence in (a-c) marked by the double arrows. In these cases, the sweep rate is set as $2\pi\times11.4$ MHz/ms.}
	\label{density-dependent oscillation}
\end{figure*}
\textbf{Ergodicity breaking from Rydberg atom clusters.}
The oscillation effect observed in Fig.~\ref{phase diagram}(e) and (f) corresponds to the broken ergodicity, which is induced by inhomogeneous Rydberg excitation in the thermal gas, resulting in spatial clusters which violate translational invariance~\cite{reichl_modern_2016,letscher_bistability_2017}. Here we model the non-ergodic dynamics by a Lindblad master equation $\dot{\rho}={\cal L}(\rho)$, where the generator $ {\cal L}(\cdot)=-i[\hat{H},(\cdot)]+\gamma\sum_{j}( J_j(\cdot)J_j^\dagger-\frac{1}{2}\{J_j^\dagger J_j,(\cdot)\})$ with jump operator $J_j=|g_j\rangle\langle r_j|$~\cite{lee2012collective,carr2013nonequilibrium}, and Hamiltonian $\hat{H}$,
\begin{eqnarray} \label{Hami}
\hat{H}&&=
\sum_{j} \left[-\Delta(t) \hat{n}_j+\Omega(t)\hat{\sigma}_j^x\right] +\sum_{j<k}{V}_{jk}\hat{n}_j\hat{n}_k,
\end{eqnarray}
where ${V}_{jk}=C_6/|\mathbf{R}_j-\mathbf{R}_k|^6$ ($C_6$ to be the state-dependent dispersion coefficient) represents the van der Waals interaction between two Rydberg atoms locating at $\mathbf{R}_j$ and $\mathbf{R}_k$. Operator $\hat{\sigma}_j^x=(|r_j\rangle\langle g_j|+|g_j\rangle\langle r_j|)/2$ flips the atomic state and $\hat{n}_j=|r_j\rangle\langle r_j|$ is Rydberg density operator of the $j$-th atom. In the experiment a large fraction of atoms is excited. We consider  up to $10^3$  atoms that are randomly distributed in space such that spatial configurations are largely explored. We numerically solve the many-body dynamics with the discrete truncated-Wigner method~\cite{schachenmayer_many-body_2015,singh_driven-dissipative_2022}, which is suitable for dealing with interacting many atom systems ($N\gg 1$). 

We show mean values $n_r = \sum_j \langle \hat{n}_j\rangle$ of the Rydberg population by sweeping the laser detuning in Fig.~\ref{against scan rate}(a), which exhibit a similar profile with the optical transmission (Fig.~\ref{phase diagram}(b)). The high level of similarity permits us to gain insights from the simulation into the observed dynamics. In Fig.~\ref{against scan rate}(b1)-(b2), examples of the Rydberg population are shown when detuning is frozen at $\Delta/\gamma  = -2.35$ (marked by the vertical line in Fig.~\ref{against scan rate}(a)). It is found that dynamics of various atoms are largely different in an initial transient period, such that the underlying trajectories explore the phase space. At a later stage ($\gamma t> 50$), some atoms reach a steady population and becomes dynamically inactive, see Fig.~\ref{against scan rate}(b1). A large fraction of the atoms, on the other hand, is dynamically active, as shown in Fig.~\ref{against scan rate}(b2). 

Surprisingly, the overall oscillations of different Rydberg atoms are synchronized at a later time, though each oscillations are at different amplitudes and frequencies. In Fig.~\ref{against scan rate}(c), we plot $n_r$ corresponding to the data shown in Fig.~\ref{against scan rate}(b1)-(b3). After a transient period (A in Fig.~\ref{against scan rate}(c)), the mean population enters an oscillation phase. Importantly we find that oscillations of different atoms are strongly synchronized, and centered at a single frequency (inset of Fig.~\ref{against scan rate}(c)). Such synchronized oscillation reproduces the essential character observed in our experiment, as shown in Fig.~\ref{phase diagram}(e).  We shall point out that the EIT setting allows to monitor transmission continuously, i.e. probing many-body dynamics without destroying Rydberg populations.

A closer look shows that the active atoms form \textit{spatial clusters}. Here the Rydberg level is shifted by the attractive van der Waals interaction, while Rydberg excitation happens under the antiblockade condition where the red detuned laser compensates the interaction~\cite{Facilitation_Marcuzzi_2017, lesanovsky_out--equilibrium_2014, Wintermantel_2021}. Features of the Rydberg distribution can be analyzed by Hopkins statistic~\cite{clusters_Hopkins_2004}, which gives a figure of merit of cluster tendency of Rydberg atom spatial distribution. Our numerical calculations show that the Hopkins statistic is larger than 0.5 in the limit cycle phase, strongly suggesting formation of spatial clusters. More details of the analysis can be found in \textbf{SM}.

\textbf{Stability of the ergodic breaking phase.}
The non-ergodic dynamics is stable, against temperature and density.  In our experiment, atomic densities can be varied by temperature of the gas. Under different temperatures $T=45.2^{\circ} \text{C},44.3^{\circ} \text{C}\text{, and  }43.4^{\circ} \text{C}$, we measure the probe transmission against $\Omega_{p}^{2}$ by sweeping $\Delta_{c}$ with a rate of $2\pi\times11.4$ MHz/ms. As shown in Fig.~\ref{density-dependent oscillation} (a), there are obvious dips in the oscillation region (marked by the yellow lines). The characteristic lines of these oscillations are gradually
faded out  when we decrease the atomic density, see Fig.\ref{density-dependent oscillation}(a)-Fig.\ref{density-dependent oscillation}(c) for comparison. Due to the resonance condition of probe field, the photoelectric signal under lower density is much larger than higher density. This effect causes a lower criticality of phase transition at relatively lower atomic densities.

We also extract the probe transmission data marked by the double arrows shown in Fig.~\ref{density-dependent oscillation}
(d-f). In Fig.~\ref{density-dependent oscillation} (d), A, B and C indicate dips in the oscillations when sweeping the detuning. The dips are gradually disappeared when we decrease the atomic density, as seen in Fig.~\ref{density-dependent oscillation}
(e-f).  The reason is that average distances between Rydberg atoms are larger if atomic densities are low. This results to weaker Rydberg atom interactions, such that dissipation overrides the dynamics. If we decrease the atomic density further, the phase transition would
disappear completely and the probe transmission spectrum becomes smooth. The obtained density-dependent oscillations behavior manifests strong many-body characters in the non-equilibrium dynamics.



\textbf{Summary.} We have studied the non-ergodic dynamics of non-equilibrium phase transition in a strongly interacting Rydberg gases. When the system approaches to the criticality, the Rydberg population is bifurcated into E- and NE-phases. In the vicinity of the critical point, we have shown the Rydberg population is periodically oscillated for a long period of time. We have also observed the density-dependent oscillation, revealing the correspondence of a many-body effect. The ergodicity breaking observed in our experiment is explained with the formation of Rydberg clusters. Early works have revealed the importance of inhomogeneous Rydberg excitation in the study of Rydberg soft matter, such as aggregates~\cite{garttner_dynamic_2013,lesanovsky_out--equilibrium_2014,schempp2014full,urvoy_strongly_2015}, bistability~\cite{letscher_bistability_2017}, and self-organization ~\cite{helmrich2020signatures,Ding_Phase_2020}. Our study shows that clusters of Rydberg atoms triggers non-equilibrium, non-ergodic many-body dynamics, despite the strong dissipation.  Experimentally exploring broken ergodicity in a driven-dissipative Rydberg gas platform will expand the category of ergodicity of complex matter and non-equilibrium phenomenon~\citep{serbyn2021quantum, Keesling_2019}, uncover the relation between the dissipation and the ergodicity~\cite{Buca_Non_2019}, and find quantum technological applications~\cite{ripka_room-temperature_2018,verstraete_quantum_2009}.

\textbf{Acknowledgements.} We thank Igor Lesanovsky, Federico Carollo, and Hannes Busche for useful discussions. We acknowledge funding from National Key R\&D Program of China (2022YFA1404002), the National Natural Science Foundation of China (Grant Nos. U20A20218, 61525504, 61722510, 61435011, 11904104, 12274131), the Youth Innovation Promotion Association of Chinese Academy of Sciences under Grant No. 2018490, the Major Science and Technology Projects in Anhui Province (Grant No. 202203a13010001), the Shanghai Pujiang Program (21PJ1402500), EPSRC through grant agreements EP/V030280/1 (``Quantum optics using Rydberg polaritons''), EP/W015641/1 (``Simulating ultracold quantum chemistry at conical intersections''), as well as, DSTL, and Durham University.).

\bibliography{ref}

\end{document}


\begin{center}
{\Large Supplementary information for}
\end{center}
\begin{center}
 {\Large \bf Ergodicity breaking from Rydberg excitation clusters in a driven-dissipative many-body system}
\end{center}
\begin{center}
{\large Dong-Sheng Ding, Zhengyang Bai, Zong-Kai Liu, Bao-Sen Shi, Guang-Can Guo, Weibin Li and C. Stuart Adams}
\end{center}
\vspace{10mm}
\setcounter{secnumdepth}{2}
\setcounter{equation}{0}
\setcounter{figure}{0}
\setcounter{table}{0}
\setcounter{page}{1}
\makeatletter

\renewcommand{\theequation}{S\arabic{equation}}
\renewcommand{\thefigure}{S\arabic{figure}}
\renewcommand{\bibnumfmt}[1]{[S#1]}
\renewcommand{\citenumfont}[1]{S#1}

In this supplementary material we give more details on the experiment and theoretical analysis. 
\section{Experimental details}

\subsection{Experimental setup}

\begin{figure}
	\centering
	\includegraphics[width=0.75\textwidth]{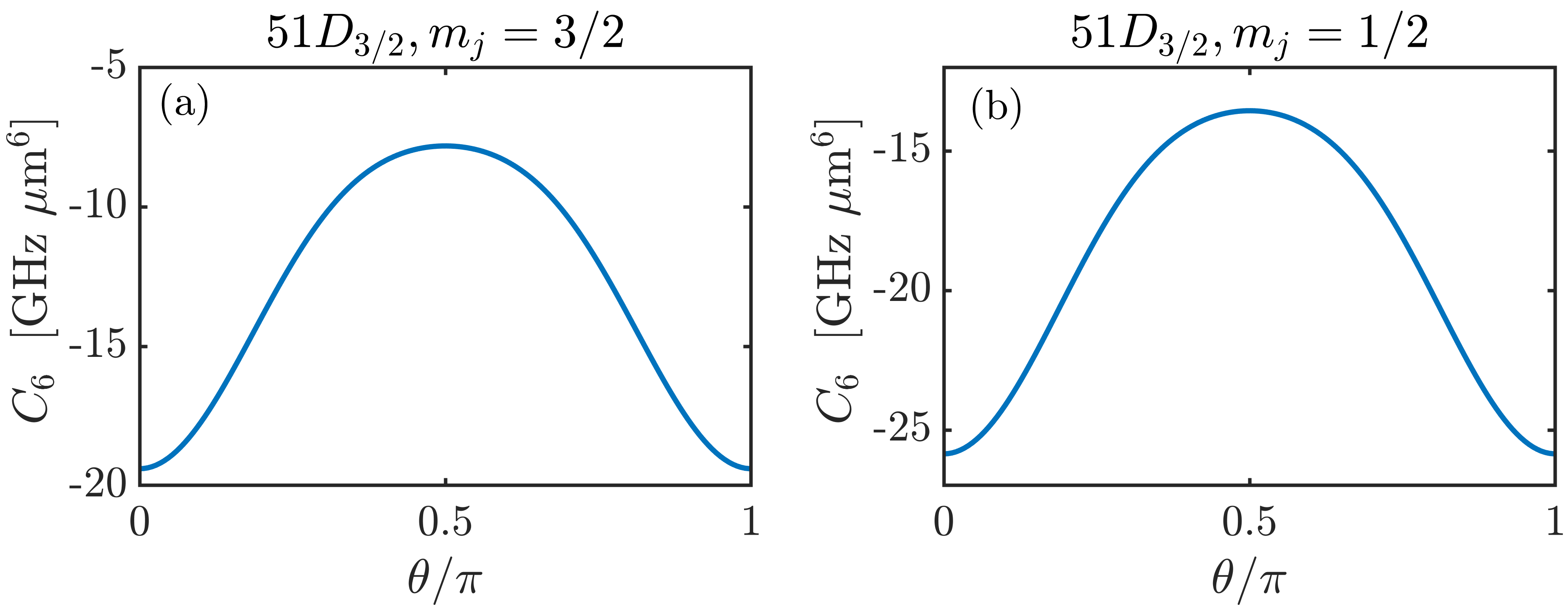}\\
	\caption{\footnotesize {\rm Angular dependent dispersion coefficient $\tilde{C}_6(\theta)$ in states $|51D_{3/2}\rangle$}.  Mean value $C_6$ is used in the numerical simulations.}\label{C6}
\end{figure}

In our experiment, three energy levels ($|g\rangle=\left| 5S_{1/2},F=2 \right\rangle$, $|e\rangle=\left| 5P_{1/2},F'=3 \right\rangle$, $|r\rangle=\left| 51D_{3/2} \right\rangle$) of Rubidium-85 are considered. The interaction between $\left| 51D_{3/2} \right\rangle$ is attractive with dispersive coefficient $C_6<0$ [see Fig.~\ref{C6}]. The experiment is carried out in a 10-centimeters-long glass cell with saturated pressure. We control the temperature to be $45^\circ \text{C}$ (corresponding to the atomic densities of $9.0\times10^{10} $ $\mathrm{cm^{-3}}$) in Figure 2 of the main text and $T=45.2^{\circ} \text{C},44.3^{\circ}\text{C},43.4^{\circ}\text{C}$ (corresponding to the atomic densities of $9.52,8.78,8.11\times10^{10}$ $\mathrm{cm^{-3}}$, respectively) in Figure 4 of the main text. Two laser beams, the probe light and the coupling light, couple the three level atoms and form a two photon excitation process via electromagnetically induced transparency (EIT), where the probe light (Toptica DL pro @795 nm) couples energy levels $\left| g \right\rangle \leftrightarrow \left| e \right\rangle$ and the coupling light (Toptica TA-SHG pro @480 nm) couples energy levels $\left| e \right\rangle \leftrightarrow \left| r \right\rangle$. Before the 795 nm laser enters the rubidium vapor cell, it is split into the probe light (with waist 1100 $\mathrm{\mu m}$) and an identical reference light parallel to the probe light by a beam displacer. Both of the probe light and the reference beam transmit through the rubidium vapor cell. While the coupling light (with waist 1000 $\mathrm{\mu m}$) counter-propagates with the probe light and cover the probe light but not the reference light to reduce the Doppler effect. Finally, the difference between the transmission signals of the probe and the reference beams are detected by a differential detector (Thorlabs PB450A). The transmission spectral therefore are obtained via scanning the detuning of the coupling light $\Delta_c$ when the frequency of the probe light is locked to the atomic transition $\left| g \right\rangle \leftrightarrow \left| e \right\rangle $. The phase diagrams (Figures 2(a), 4(a-c) in the main text) are obtained via first sweeping the detuning $\Delta_c$ with rate $2\pi\times6$ MHz/ms and then tuning the Rabi frequency of the probe light $\Omega_p$ by changing its power and again sweeping the detuning $\Delta_c$ for the next Rabi frequency $\Omega_p$.

\section{Theoretical Model of the many-body driven-dissipative dynamics}

\subsection{Master equation}
The dynamics of the Rydberg atom can be well described by an effective two-level model~\cite{Carr_Nonequilibrium_2013,Ding_Phase_2020}. We consider an ensemble of $N$ two-level atoms. The electronic ground state $|g\rangle$ is coupled to excited state $|r\rangle$ by laser fields with an effective, time-dependent Rabi frequency $\Omega(t)$ and detuning $\Delta(t)$. Atoms in state $|r\rangle$ interact
strongly through van der waals interaction ${V}_{jk}=C_6/|\mathbf{R}_j-\mathbf{R}_k|^6$ with $C_6$ and $\mathbf{R}_{j(k)}$ the dispersion coefficient and location of the $j(k)$-th atom.  Hamiltonian $\hat{H} of the many-body system reads (\hbar\equiv1)$,
\begin{eqnarray} \label{Hami}
\hat{H}&&=
\sum_{j} \left[-\Delta(t) \hat{n}_j+\Omega(t)\hat{\sigma}_j^x\right] +\sum_{j<k}{V}_{jk}\hat{n}_j\hat{n}_k,
\end{eqnarray}
where $\hat{\sigma}_j^x=(|r_j\rangle\langle g_j|+|g_j\rangle\langle r_j|)/2$ flips the atomic state and $\hat{n}_j=|r_j\rangle\langle r_j|$ is the projection operator of the excited state. 
Including the dissipation, dynamics of the system density matrix $\rho$ is modelled by a Lindblad master equation,
\begin{eqnarray}\label{master}
\dot{\rho}(t)={\cal L}\rho(t),
\end{eqnarray}
where ${\cal L}(\cdot)=-i[\hat{H},(\cdot)]+\gamma\sum_{j}( J_j(\cdot)J_j^\dagger-\frac{1}{2}\{J_j^\dagger J_j,(\cdot)\})$ with $J_j=|g_j\rangle\langle r_j|$ is the jump operator~\cite{lee2012collective, Carr_Nonequilibrium_2013}.

\subsection{Mean field approach}
For small systems (i.e., about 10 atoms), the quantum master equation can be solved numerically.
However, the Hilbert space of the Hamiltonian grows with $2^N$, while the dimension of the density matrix is $2^{2N}$. The computational complexity prevents us from numerically solving the many-body problems when $N > 10$ in typical computers. Due to the dissipation, many-body correlations may be weak, such that we could employ approximations, such as the mean-field (MF) theory and truncated discrete Wigner method to simulate the dynamics.

In the MF approach, the many-body density matrix $\rho$ is decoupled into tensor products of individual ones $\hat{\rho} \approx \Pi_i$ $\hat{\rho}_i$. This decoupling essentially ignores correlations between different sites~\cite{diehl2010dynamical}.
This is a good approximation in a three-dimensional system with large number of atoms (typically $10^3\sim10^4$ Rydberg atoms are prepared in the experiment).
In the MF calculation, mean values of spin operators $s^{\mu}_j=\langle\hat{\sigma}^{\mu}_j\rangle~(\mu=x, y, z)$ are calculated, whose dynamics is governed by the following equations of motion,
\begin{subequations}\label{spinmeanfield}
\begin{eqnarray}
&&\frac{ds^x_j}{dt}=-\Delta s^y_j-\frac{\gamma}{2} s^x_j+\sum_{j<k}{V}_{jk}s^y_jn_{r}^k,\\
&&\frac{ds^y_j}{dt}=\Delta s^x_j-\frac{\gamma}{2} s^y_j-\Omega s^z_j-\sum_{j<k}{V}_{jk}s^x_jn_{r}^k,\\
&&\frac{ds^z_j}{dt}=\Omega s^y_j+\frac{\gamma}{2}(1-2s^z_j),
\end{eqnarray}
\end{subequations}
where $n_{r}^j=0.5-s^z_j$ represent the Rydberg population on site $\mathbf{R}_j$. 

\subsection{Mean field phase diagram}
\begin{figure}
	\centering
	\includegraphics[width=0.75\textwidth]{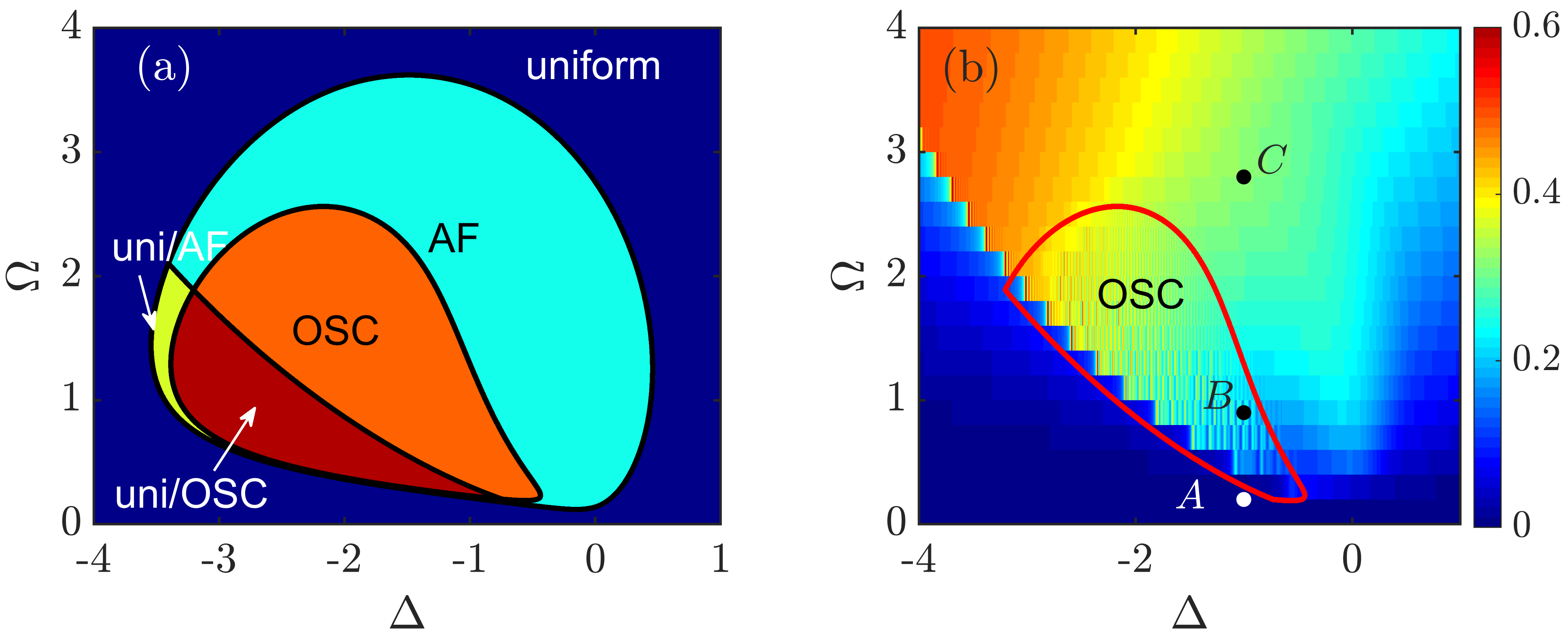}\\
	\caption{\footnotesize {\rm MF phase diagrams.} (a) The phase diagram is mainly occupied by the UNI, AF, and OSC phase.  (b) The many-body phases could be detected by sweeping detuning $\Delta$ from red to blue side linearly (i.e., $\Delta(t)=\Delta_0+a\times t$). The Rydberg population $n_r(t)$  in the $\Delta-\Omega$ plane is sketched. The oscillatory phase can be identified by Rydberg excitation. In our calculation, we select $V_{AB}=-8$, $\gamma=0.5$ and $a=0.01$.}\label{MF_Phase_diagram}
\end{figure}

The limit cycle and uniform phase can already be revealed from the MF theory. We first investigate a bipartite (i.e. two sies labelled by $A$ and $B$) subspace that represents the minimal model of our setting. In this case, Eq.~(\ref{spinmeanfield}) is simplified to the following coupled equations of the $A-B$ subspace,
\begin{subequations}\label{eq:AB}
\begin{eqnarray}
&&\frac{ds^x_A}{dt}=-\Delta s^y_A-\frac{\gamma}{2} s^x_A+{V}_{AB}s^y_A n_{r}^B,
\\
&&\frac{ds^y_A}{dt}=\Delta s^x_A-\frac{\gamma}{2} s^y_A-\Omega s^z_A-{V}_{AB}s^x_An_{r}^B,\\
&&\frac{ds^z_A}{dt}=\Omega s^y_A+\frac{\gamma}{2}(1-2s^z_A),
\end{eqnarray}
\end{subequations}
Equations for $B$ site can be obtained by swapping index $A$ and $B$ in Eq.~(\ref{eq:AB}). The fixed points of Eqs.~(\ref{eq:AB}) can be found by setting $\dot{s}^x_{A(B)}=\dot{s}^y_{A(B)}=\dot{s}^z_{A(B)}=0$. Two types of fixed points are found: a uniform phase (i.e., $s_z^A=s_z^B$) correspond to spatially homogeneous excitation, and a nonuniform one (i.e., $s_z^A\neq s_z^B$) could contribute to both the antiferromagnetic  and oscillatory phases.
We further analyze the linear stability of the fixed points by calculating eigenvalues $\lambda_j$ of the Jacobian matrix of Eqns. (\ref{eq:AB})~\cite{Strogatz2015Nonlinear}. If the real parts of all eigenvalues are negative, the corresponding solution is stable; otherwise, it is unstable. MF phase diagrams of the bipartite subspace are shown in Fig.~(\ref{MF_Phase_diagram})(a). We find that the system is dominated by three cases: uniform (UNI), antiferromagnetic (AF), and oscillatory (OSC). The latter two correspond to nonuniform phases. 

Dynamically one can probe the many-body MF phases by sweeping the laser detuning $\Delta$ from red to blue side linearly (i.e., $\Delta(t)=\Delta_0+a\times t$). When the detuning $\Delta$ approaches the resonance, the population of Rydberg atoms bifurcates from the UNI phase to  oscillatory phases, which is found when $\Omega>0.4$  [Fig.~(\ref{MF_Phase_diagram})(b)].

\begin{figure}
	\centering
	\includegraphics[width=0.95\textwidth]{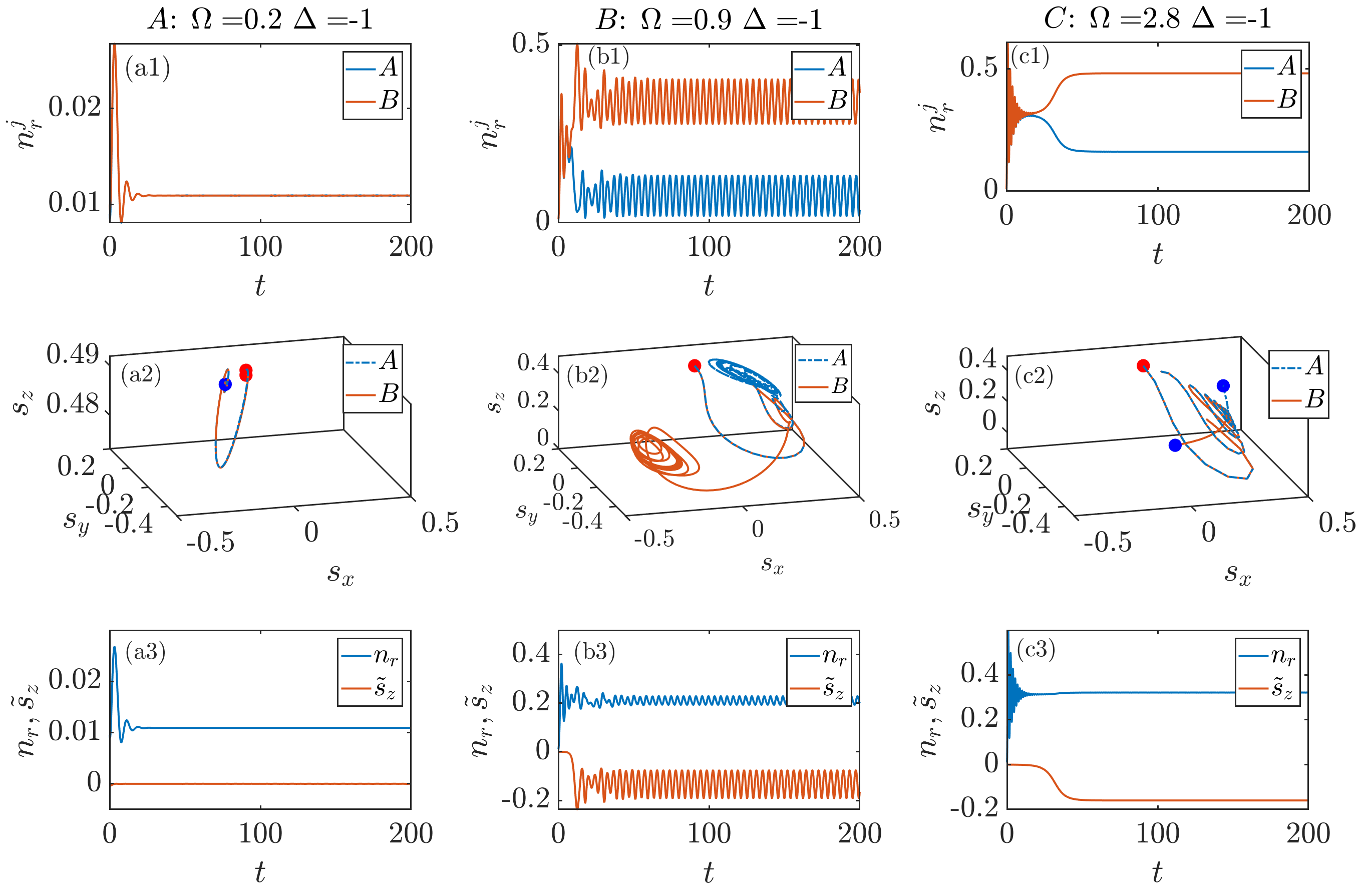}\\
	\caption{\footnotesize {\bf Excitation of different phases.} (a1) Rydberg populations $n_r^j$ $(j=A, B)$ as a function $t$ with $\Delta=-1$ and $\Omega=0.2$. The system is occupied by UNI phase. The corresponding phase space trajectories are plotted in panel (a2). The system starts from red dot and decay to the fixed point (blue dot). (a3) shows population difference $\tilde{s}_z=s_z^A-s_z^B$ and the average Rydberg population $n_r$. For UNI phase, $\tilde{s}_z=0$. (b1) Evolution of $n_r^j$ with $\Delta=-1$ and $\Omega=0.9$. The system is in the OSC phase. Their phase space trajectories begin to oscillate persistently, and give rise to limit cycles [see panel (b2)]. (b3) $\tilde{s}_z$ and $n_r$ are also non-stationary. (c1) The system is in the AF phase when increasing $\Omega$ to $2.8$. (c2) The phase space trajectories decay to two different fixed points (blue dot). (c3) The order parameter $\tilde{s}_z$ is non-zero in AF phase. The panel (a), (b) and (c)  correspond to points $A$, $B$, and $C$ in Fig.~\ref{MF_Phase_diagram}(b).}\label{Two_atom_dynamics}
\end{figure}

Fig.~\ref{Two_atom_dynamics} shows the dynamics of the  Rydberg populations with different $\Omega$. When $\Omega$ is small, the steady state is the UNI phase [Fig.~\ref{Two_atom_dynamics}(a1)] where  excitation of atoms are identical, then changes to the OSC phase [Fig.~\ref{Two_atom_dynamics}(b1)] and then to the AF phase by increasing $\Omega$ [Fig.~\ref{Two_atom_dynamics}(c1)]. For UNI phase, atoms decay to a single fixed point (blue dot) [Fig.~\ref{Two_atom_dynamics}(a2)]. These unstable nonuniform fixed points lead to limit cycles [Fig.~\ref{Two_atom_dynamics}(b2)], in which the Rydberg population oscillates periodically in time [Fig.~\ref{Two_atom_dynamics}(b3)]. Further increasing $\Omega$, the nonuniform phase becomes stable which means the system shows an antiferromagnetic pattern [Fig.~\ref{Two_atom_dynamics}(c1)]. Their phase space trajectories can decay to two different fixed points [Fig.~\ref{Two_atom_dynamics}(c2)]. In both the uniform and antiferromagnetic phase, the population dynamics is stationary, which is in sharp contrast to the population dynamics found in the oscillatory phase. 

\subsection{Discrete truncated Wigner approximation}

To simulate the experimental situation, we consider an ensemble of $N$ atoms that are randomly distributed in space [see Fig. 3(b3) in the main text]. The number of Rydberg excitation is large (i.e., $10^3\sim10^4$) in the experiment. To efficiently characterize the quantum many-body behaviour of such system,  we apply the discrete truncated Wigner approximation (DTWA; beyond MF phase space method), where the interaction $\hat{n}_j\otimes\sum_{k}\hat{n}_k$ is replaced with the MF term $\hat{n}_j\sum_{k}\langle\hat{n}_k\rangle$, and the quantum fluctuations are incorporated in initial states by a Wigner distribution~\cite{Schachenmayer_DTWA_2015}.

We have shown the dynamical evolution of the Rydberg population in Fig.~3(b1)-(b2) in the main text. Based on the dynamical behavior, the atomic system can be distinguished into two components: Most of the atoms reaches a steady population and becomes dynamically inactive ($N_{\rm in}$ atoms labelled with small gray circles), and another one is non-stationary ($N_{\rm active}$ atoms labelled with large red circles). We label the non-stationary solutions as the active atoms.
The active atoms form spatial clusters in the Rydberg gas and give rise to the oscillatory phase into the ensemble.

\begin{figure}
	\centering
	\includegraphics[width=1\textwidth]{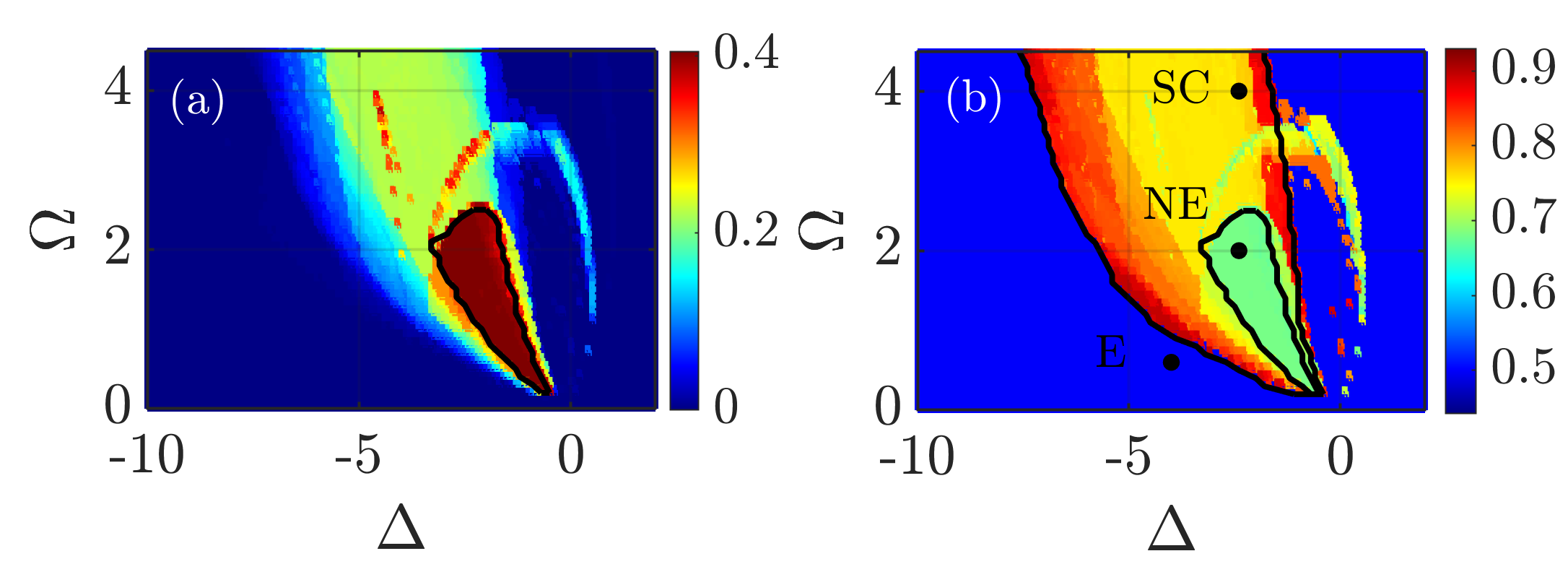}\\
	\caption{\footnotesize  {\rm DTWA phase diagram and Hopkins statistic.}  We use the different order parameters to sketch the DTWA phase diagram in $\Omega-\Delta$ space. (a) The ratio $N_{\rm active}/N$ is used as order parameter. (b) The Hopkins $H$ is calculated to identify the formation of Rydberg cluster, where $H\simeq 0.5$ indicates homogeneous distribution and $H> 0.5$ means the Rydberg clusters are forming.  Based on the magnitude of $H$, the system can be classified into E-, NE-, and SC- phases. 
}\label{phase_diagram}
\end{figure}

For random gas, we can use the ratio $N_{\rm active}/N$ to character active fraction of the system. Here the ratio $N_{\rm active}/N\rightarrow1$ means that most of the atoms is dynamically active, and  $N_{\rm active}/N\rightarrow0$ denotes the system works in inactive phase. Examples of DTWA phase diagrams (with order parameter $N_{\rm active}/N$) are shown in Fig.~\ref{phase_diagram}(a). It is found that the active phase emerges in the black curve marked region where $N_{\rm active}/N>0.4$.

\subsection{Hopkins statistic}
Under strong interaction regime, active atoms form Rydberg clusters, which violates the translation invariance and induces the broken ergodicity in free space.
To identify the formation of Rydberg clusters, we calculate the Hopkins statistic~\cite{clusters_Hopkins_2004}. The strategy of Hopkins test statistics is the follows: We randomly sample $M$ test points  [i.e., ${\mathbf X_t}= (x^t_j, y^t_j, z^t_j)$, $j=1,2,..M$] in our simulation space.  As demonstrated above, the position of active atom can be identified with ${\mathbf X_a}= (x^a_j, y^a_j, z^a_j)$, ($j=1,2,..N_{\rm active}$). Two types of distances can be defined: (i) The minimum distance $u_j$ is found from ${\mathbf X_t}$ to its nearest site in active ensemble  ${\mathbf X_a}$; (ii) We can also calculate the minimum distance $w_j$ from a randomly selected site in ${\mathbf X_a}$ to its nearest neighbor ($M$ out of the $N$ sites are marked randomly).
The Hopkins statistic in our scheme is defined as,
\begin{eqnarray} \label{NN}
H=\frac{\sum_{j=1}^{M}u_j}{\sum_{j=1}^{M}u_j+\sum_{j=1}^{M}w_j}.
\end{eqnarray}
This statistic compares the nearest-neighbor distribution of randomly selected locations to that for the randomly selected sites.  $H=0.5$ reflects the random distribution of atoms, and the Rydberg atoms are aggregated or clustered when $H>0.5$~\cite{clusters_Hopkins_2004}.

The Hopkins statistic of the atoms are shown in Fig.~\ref{phase_diagram}(c). It is clear to observe that: (i) In the ergodic phase, $H\simeq 0.5$ and the atomic ensemble is homogeneous; (ii) In the non-ergodic/oscillatory phase (marked by the black curve), $H\simeq 0.7$. The latter regime  perfectly overlaps with the dynamically active phase, in which the Rydberg clusters form. (iii) In a stationary cluster (SC) phase, $H\simeq 0.9$. In this regime, the number of active atom decreases, but most of inactive atoms decay to different fixed points. It results in multistable stationary phases~\cite{Carr_Nonequilibrium_2013,Ding_Phase_2020}.

%
%
\begin{figure}
	\centering
	\includegraphics[width=1\textwidth]{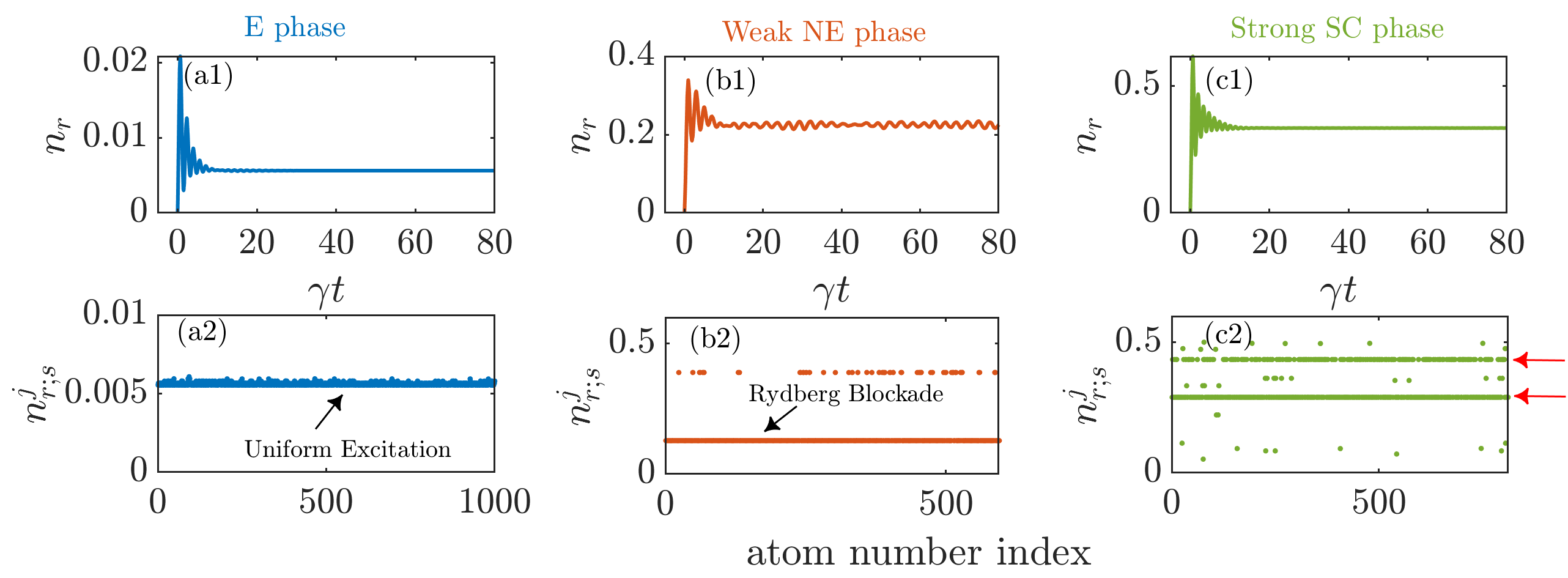}\\
	\caption{\footnotesize {\bf Dynamics of the atom ensemble.} (a1) the mean values of the Rydberg population $n_r$ as a function $t$ with $\Omega = 0.6$ and $\Delta = -4$. The system is in the E phase. We plot fixed points of corresponding inactive atoms in panel (a2). All fixed points are almost identical. (b1) $n_r^j$ varies with $t$ for $\Omega = 2$ and $\Delta = -2.4$.  The system is in the NE regime. In panel (b2), we can find that most of inactive atoms are blockaded, i.e. with very small Rydberg population. 
(c1)-(c2) The system is in the SC phase when increasing $\Omega$ to $4$. At later time, the average Rydberg population is stationary. Different atoms occupy very different Rydberg populations. Here a large fraction of the fixed points decay to two different values (highlighted by two red arrows). The panel (a), (b) and (c)  correspond to  E-, NE-, and SC- phases in Fig.~\ref{phase_diagram}(c). }\label{phase_dyn}
\end{figure}
We present three examples of the dynamical evolution in Fig.~\ref{phase_dyn}, covering the E-, NE-, and SC-phases.
In E phase, the mean values of the Rydberg population $n_r$ is weak [Fig.~\ref{phase_dyn}(a1)]. All atoms evolve into uniform fixed points with weak fluctuations [Fig.~\ref{phase_dyn}(a2)]. In NE phase, a large fraction of the atom is active dynamically. At the same time, due to the strong interaction, most of inactive atoms are blockaded and play a minor role in contributing to the Rydberg population, Here $n_r$ oscillates periodically in time [Fig.~\ref{phase_dyn}(b1)-(b2)]. In the SC phase, the number of active atoms decreases. Although clusters are found in this phase, the overall dynamics is time independent, as the number of active atoms is small.

\bibliography{References}